\documentstyle[prl,multicol,aps,epsf]{revtex}
\begin{document}
\epsfverbosetrue
\title{Fundamental and vortex solitons in a two-dimensional optical lattice}
\author{Jianke Yang$^1$ and Ziad H. Musslimani$^2$}
\address{$^1$Department of Mathematics and Statistics, University of Vermont,
Burlington, VT 05401\\
$^2$ Department of Applied Mathematics,
University of Colorado, Campus Box 526, Boulder, CO 80309-0526}
\maketitle

\begin{abstract}
Fundamental and vortex solitons in a two-dimensional optically induced
waveguide array are reported. In the strong localization regime, the
fundamental soliton is largely confined to one lattice site, while the vortex
state comprises of four fundamental modes superimposed in a square
configuration with a phase structure that is topologically equivalent to
the conventional vortex. However, in the weak localization regime, 
both the fundamental and vortex solitons spread over many lattice sites. 
We further show that fundamental and
vortex solitons are stable against small perturbations in the strong
localization regime. {\em OCIS codes: 190.0190.}
\end{abstract}

\begin{multicols}{2}
\narrowtext
Discrete optical spatial solitons are localized modes of nonlinear
waveguide arrays that form when discrete diffraction is balanced by
nonlinearity\cite{LDK}.
When a low intensity beam is focused into a waveguide array, the propagating
field spreads over many sites (due to optical tunneling) exhibiting a typical
discrete diffraction pattern with the intensity mainly concentrated
in the outer lobes \cite{Eis1}. 
However, at sufficiently high power, the beam self-traps
to form a localized state.

Discrete solitons in waveguide arrays were first predicted to
exists as solutions to the discrete nonlinear Schr\"odinger
equation\cite{demetri} and later observed in AlGaAs waveguide 
arrays\cite{Eis1}. This experimental observation stimulated much new research 
such as diffraction management\cite{Eis2,AM}, soliton interaction and beam
steering\cite{Aceves}, and discrete solitons in a two-dimensional (2D) 
photorefractive optical lattice \cite{Nikos,fleisher1,fleisher2}.
The experiments in \cite{Nikos,fleisher1,fleisher2} are particularly
interesting since the 2D waveguide array there is formed optically, thus
it is very versatile and easily tunable. This allowed a host of
2D localization phenomena to be observed in \cite{Nikos,fleisher1,fleisher2}.
However, many questions on discrete solitons in 2D optical lattices still
remain open. For instance, stability properties of such solitons have not
been carefully studied. In addition, new localization states such as those
analogous to vortex solitons in a homogeneous waveguide have not been explored.

In this Letter, fundamental and vortex solitons in a 2D optical lattice are 
found and their stability analysed. In the strong localization regime, the 
fundamental soliton is confined largely on one lattice site with a uniform phase.
The vortex soliton comprises of four fundamental modes located at the
bottoms of the optical potential
in a square configuration with a phase structure that is topologically
equivalent to the conventional vortex. By winding around the
zero intensity position along any simple closed curve, the phase of the vortex
state acquires $2\pi$ increment.
We call this structure a {\it vortex cell}.
When the localization is weak, both fundamental solitons and vortex-cells
spread over many lattice sites.
In the strong localization regime, we show that the fundamental soliton as well as
the vortex cell are both linearly and nonlinearly stable under weak perturbations.

We start with the $2D$ nonlinear Schr\"odinger equation
\begin{equation}\label{NLS}
i\frac{\partial\psi}{\partial z}+\left(\frac{\partial^2}{\partial X^2}+
\frac{\partial^2}{\partial Y^2}\right)\psi-V\psi+|\psi|^2\psi=0\;,
\end{equation}
where $V=V_0\left(\cos^2X+\cos^2Y\right)$ is the optical lattice potential, and
$V_0$ is its intensity. Such a potential can be obtained by optically 
interfering two pairs of laser beams
\cite{fleisher2}. Transverse distances $X$ and $Y$ in (\ref{NLS}) have been
normalized by the lattice period $D$, and distance $z$ normalized by
$4n_0D^2/\pi\lambda$, where $n_0$ is the refractive index, and $\lambda$ is
the wavelength of the beam \cite{fleisher2}. For typical values $D=9\mu m$, 
$n_0=2.3$ and $\lambda=0.5\mu m$ \cite{fleisher1,fleisher2}, the unit 
distance $z$ in (\ref{NLS}) corresponds to 0.5mm in physical distance.
Without the lattice potential, solitons would suffer collapse under small
perturbations \cite{zakharov}. However, as we shall show here, optical lattices
can suppress the collapse of fundamental solitons and vortex-cells.

Eq. (\ref{NLS}) conserves two quantities: the power
$P=\int_{-\infty}^\infty \int_{-\infty}^\infty |\psi|^2 dXdY$,
and the energy $E$:
\begin{equation}
E=\int_{-\infty}^\infty \int_{-\infty}^\infty
\left\{|\nabla\psi|^2-\frac{1}{2}
|\psi|^4+V|\psi|^2 \right\}dXdY\;.
\end{equation}
We look for stationary solutions of the form
$\psi(X,Y,z)=e^{-i\mu z} u(X, Y)$ where $\mu$ being the propagation constant
of the soliton. Then $u(X,Y)$ satisfies
\begin{equation}
\frac{\partial^2 u}{\partial X^2}+
\frac{\partial^2 u}{\partial Y^2}
-Vu + |u|^2u=-\mu u\;.
\label{u}
\end{equation}
Solutions to this equation can be obtained by a Fourier-iteration method
as was suggested by Petviashvili \cite{petviashvili}.

\begin{figure}
\begin{center}
\begin{tabular}{cc}
\setlength{\epsfxsize}{5.3cm}\epsfbox{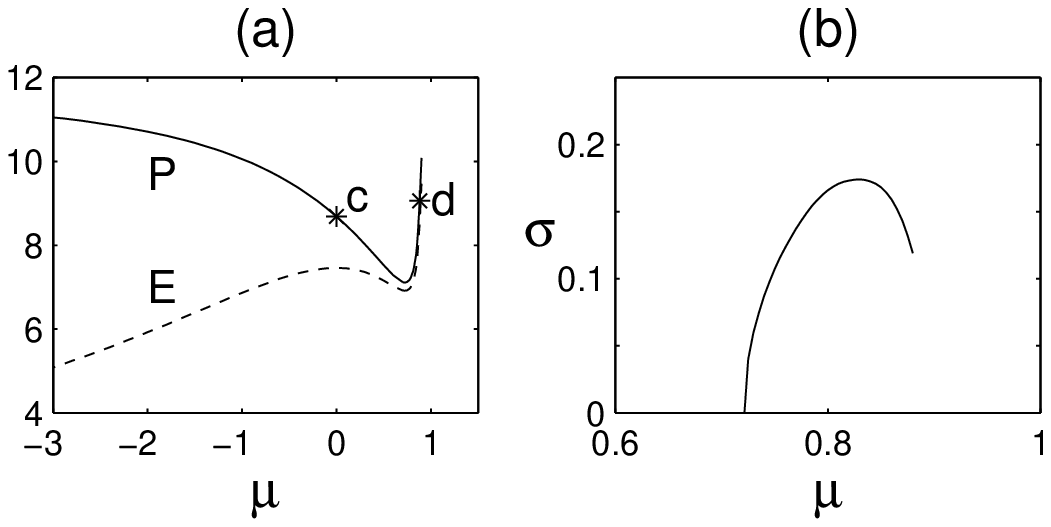}\\
\setlength{\epsfxsize}{5.3cm}\epsfbox{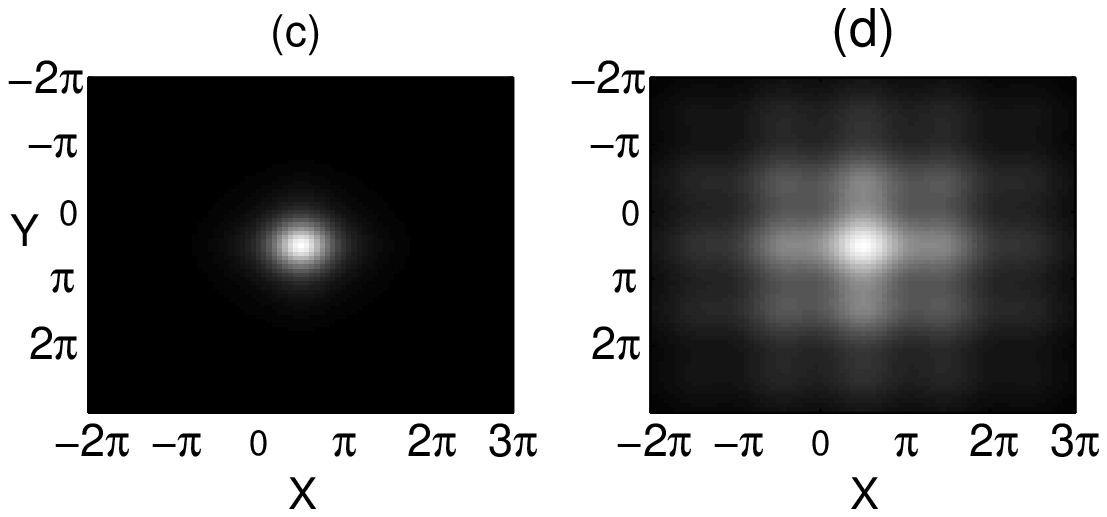}
\end{tabular}
\end{center}
\caption{(a) Normalized power $P$ and energy $E$
of fundamental solitons versus $\mu$ for $V_0=1$.
(b) unstable eigenvalues $\sigma$ of these solitons
for $V_0=1$; (c, d) profiles of fundamental solitons at $\mu=0$ and
$0.88$ ($V_0=1$) respectively.}
\vspace{-0.4cm}
\label{fundamental1}
\end{figure}

{\it Fundamental solitons.} A fundamental soliton of Eq. (\ref{u}) 
has a single main hump sitting at a bottom of the potential, say
$(X, Y)=(\pi/2, \pi/2)$. Two examples corresponding to propagation constants
$\mu=0$ and $0.88$ with $V_0=1$ are displayed
in Fig.~\ref{fundamental1}(c,d). We see that for small $\mu$
(Fig. \ref{fundamental1}c), the
beam is largely confined on one lattice site,
while at higher $\mu$ (Fig. \ref{fundamental1}d), it spreads over many lattice
sites. To quantify these solitons,
we calculate the dependence of the normalized power $P$
on the propagation constant
$\mu$ for $V_0=1$ as displayed in Fig.~\ref{fundamental1}a.
When $\mu \to -\infty$, $P$ approaches a constant 11.70.
This is apparently because in this limit, the fundamental soliton is highly
localized, thus it approaches the lattice-free fundamental-soliton state,
which has critical power $P_c\approx 11.70.$
As $\mu$ goes to a cut-off value which is approximately 0.95,
$P$ appears to go to infinity. In this limit, the fundamental state becomes
uniformly distributed in space.
Thus, this cut-off value should be the boundary of
the band gap in the linear-wave
spectrum \cite{carusotto}.
When $\mu\approx 0.72$, $P$ is minimal.
When $\mu > 0.72$, $dP/d\mu >0$.
In this region, the Vakhitov-Kolokolov (VK) theorem
suggests that the soliton is linearly unstable due to the presence of
a purely-real unstable eigenvalue $\sigma$ in the linearized equation
\cite{VK}. We have confirmed this instability by numerically
simulating the linearized version of
(\ref{NLS}) around the above soliton. The results for unstable 
eigenvalues are shown in Fig.~\ref{fundamental1}b. On the other hand, 
when $\mu < 0.72$ ($V_0=1$), the VK instability is absent, and the soliton 
state is linearly stable.

\begin{figure}
\begin{center}
\begin{tabular}{cc}
\setlength{\epsfxsize}{5.3cm} \epsfbox{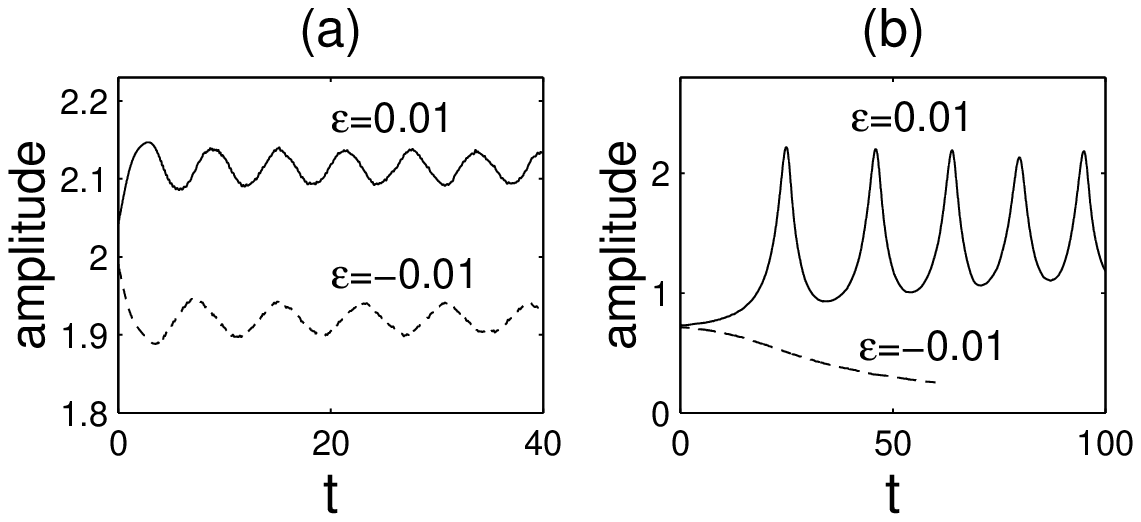}\\
\setlength{\epsfxsize}{5.3cm} \epsfbox{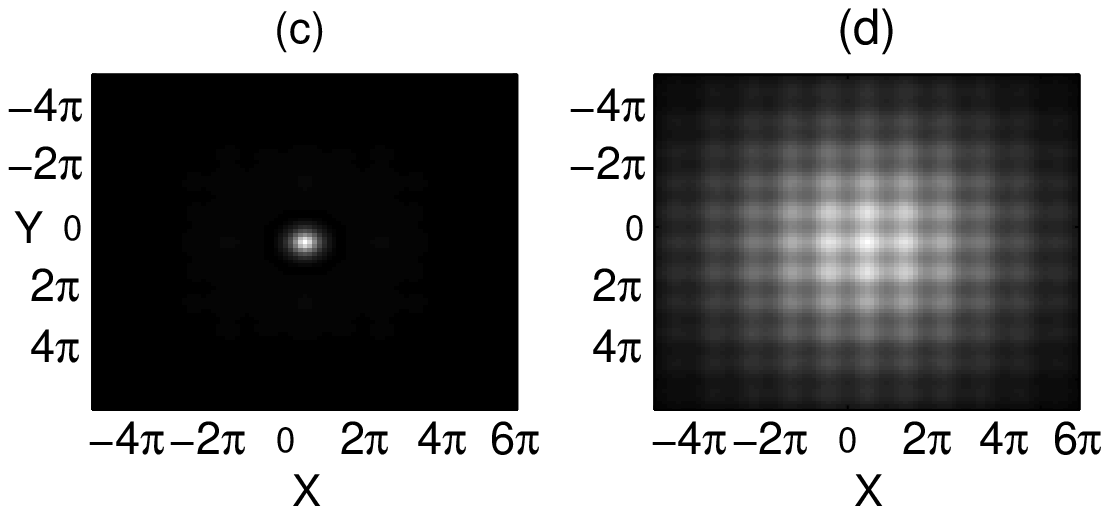}
\end{tabular}
\end{center}
\caption{Nonlinear evolutions of fundamental solitons under
perturbations (\ref{ic}) for $V_0=1$,
$\epsilon=\pm 0.01$, $u_p=1$ and $\mu=0$ (a), $0.88$ (b).
Snapshots of the soliton intensity corresponding to
the dynamics depicted in (b) for $\epsilon=0.01$, $z=80$ (c) and
$\epsilon=-0.01$, $z=60$ (d).}
\vspace{-0.3cm}
\label{fundamental2}
\end{figure}

For the two-dimensional self-focusing case,
collapse is an important issue. The above linear stability analysis
does not guarantee that the fundamental soliton will not collapse
under small perturbations. In the study of collapse, the energy $E$ plays an
important role. In the absence of the lattice potential\cite{zakharov} or when the
potential is harmonic\cite{pitaevskii}, the soliton 
collapses if its energy is negative. In case the energy is positive,
however, the soliton collapses only if it is strongly perturbed.
For the optical lattice, we have calculated the energy $E$
of fundamental solitons at various values of $\mu$ and plotted the results
in Fig. 1(a). The energy is found to be always positive.
Thus we can expect that this state
is able to withstand small perturbations without collapse.
To confirm the above expectations on stability, we numerically study
the nonlinear evolution of the fundamental soliton under
small perturbations by directly simulating Eq.~(\ref{NLS}) with initial
condition
\begin{equation} \label{ic}
\psi (X,Y,z=0)=u(X,Y)[1+\epsilon u_p(X,Y)]\;,
\end{equation}
where $\epsilon\ll 1$, and $u_p(X,Y)$ is the initial perturbation.
We first take $u_p$ to be white noise. A large number of simulations
with small $\epsilon$ and
various realizations of random-noise perturbations have been
performed, and we have found that for $V_0=1$, if
$\mu<0.72$, the fundamental soliton is indeed stable against white noise
perturbations; when $\mu>0.72$, the soliton is unstable.
To study the nonlinear evolution process, we now take $u_p=1$.
For $V_0=1$, $\mu=0$ and $\epsilon=\pm 0.01$, nonlinear evolutions
are plotted in Fig.~\ref{fundamental2}a.
We see that the perturbed soliton only oscillates weakly around the
fundamental-soliton state, meaning that the soliton is both linearly and 
nonlinarly stable. On the other hand, at $V_0=1$ and $\mu=0.88$ (where the 
soliton is linearly unstable), the dynamics is different as two scenarios are 
identified: (i) at higher input power ($\epsilon >0$),
the perturbed state relaxes into a $z$-periodic bound
state (Fig.\ref{fundamental2}b,c); (ii) at lower input power
($\epsilon <0$), the perturbed state decays into linear Bloch waves
(Fig. \ref{fundamental2}b,d) (similar scenarios can be found in \cite{pelinovsky}
for a different system). Collapse is not observed in either scenarios.

\begin{figure}
\begin{center}
\begin{tabular}{cc}
\setlength{\epsfxsize}{5.5cm} \epsfbox{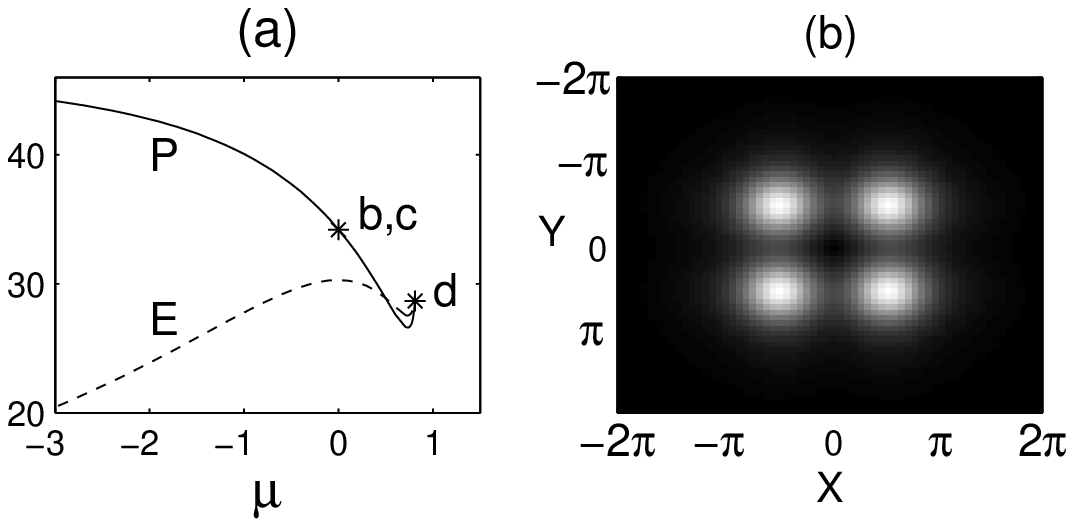}\\
\setlength{\epsfxsize}{5.5cm} \epsfbox{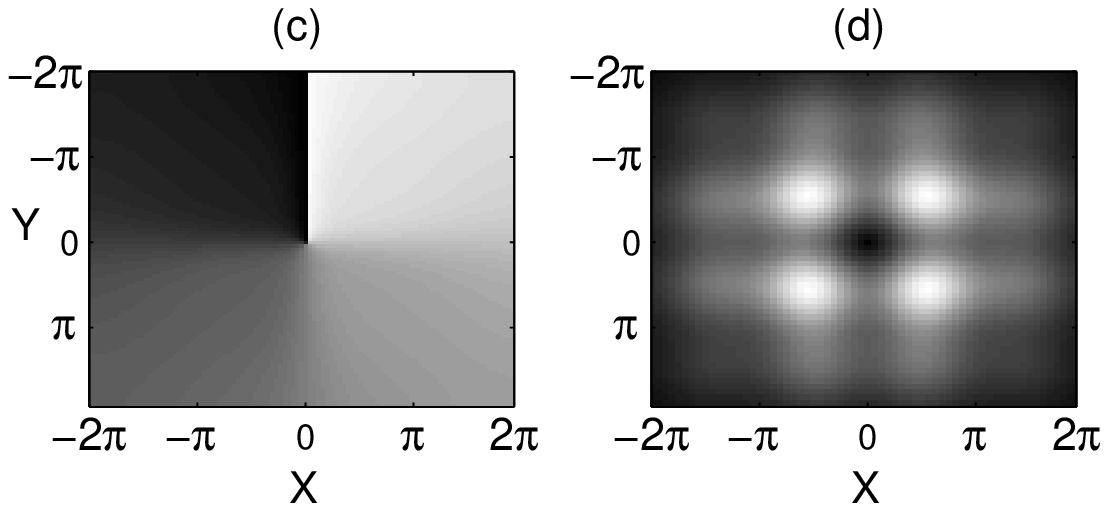}
\end{tabular}
\end{center}

\vspace{-0.1cm}
\caption{(a) Normalized power $P$ and energy $E$
of vortex cells versus $\mu$ for $V_0=1$.
(b,d) intensity plots of vortex cells with $V_0=1$, $\mu=0$ and $0.82$
respectively. (c) the phase plot of the vortex cell in (b).}

\vspace{-0.3cm}
\label{vortexfig}
\end{figure}

\begin{figure}
\begin{center}
\begin{tabular}{cc}
\setlength{\epsfxsize}{5cm} \epsfbox{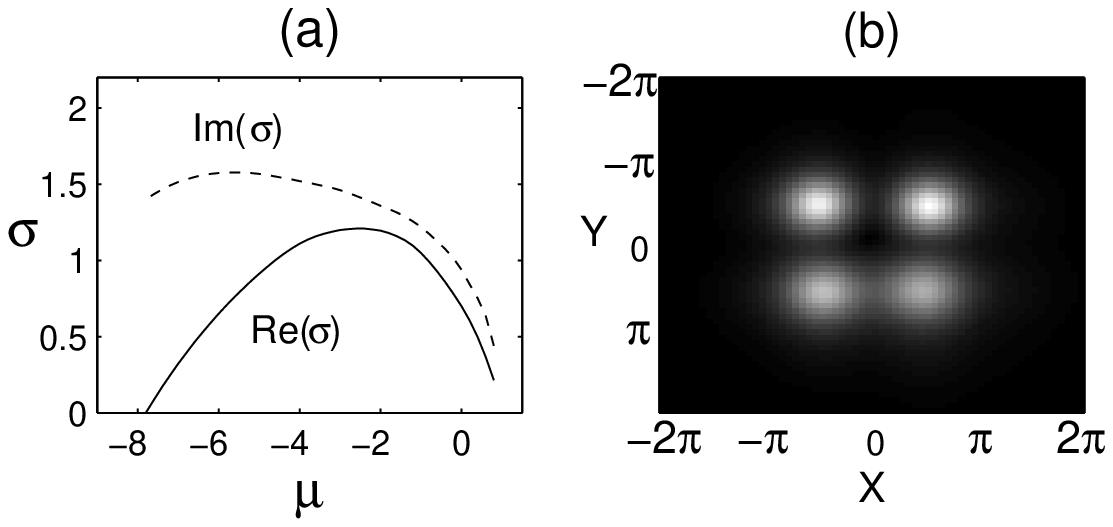}\\
\setlength{\epsfxsize}{5cm} \epsfbox{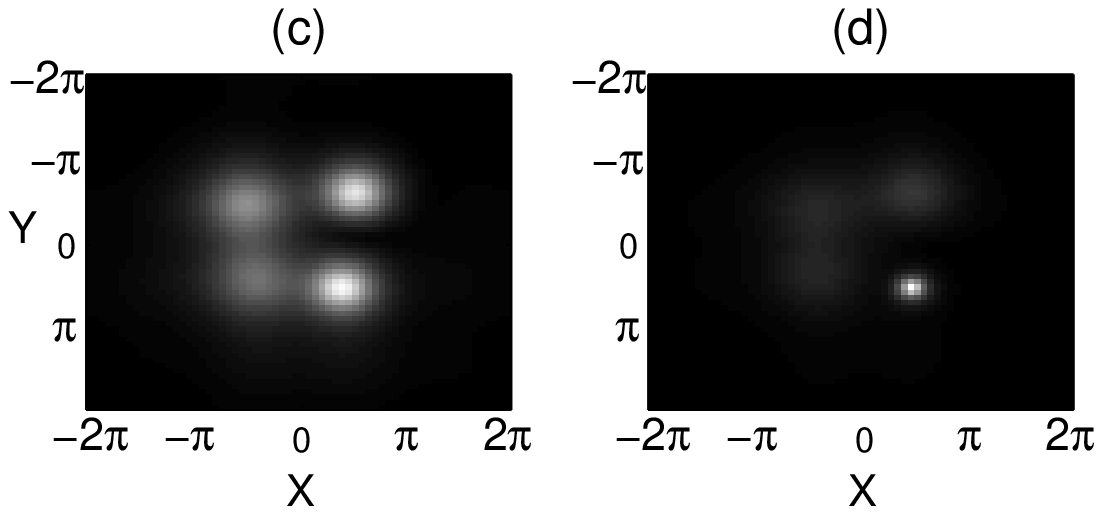}
\end{tabular}
\end{center}

\vspace{-0.3cm}
\caption{(a): Unstable eigenvalues of vortex cells with $V_0=1$;
(b,c,d) instability development of the vortex with $V_0=1$ and $\mu=0$
[see Fig.~\ref{vortexfig}(b,c)]
when it is initially amplified by 1\%;
intensity plots at $z=50, 51.5$ and 52 are shown respectively.}

\vspace{-0.3cm}
\label{evolution}
\end{figure}

{\it Vortex cell.}
In addition to the fundamental solitons, we have numerically found vortex
solitons as well. Two examples with $V_0$=1, $\mu=0$ and $0.82$ are shown in
Fig.~\ref{vortexfig} (b,c,d).
At $\mu=0$ (strong localization regime), the vortex state comprises of four
fundamental solitons superimposed in a square configuration with a phase
structure which is topologically equivalent to a conventional vortex
[see Fig.~\ref{vortexfig}b,c]. By winding around the center along any closed
curve, the phase of the vortex acquires $2\pi$ increment, thus, we name it a
vortex cell. At $\mu=0.82$ (weak localization regime), the vortex cell
spreads out to more lattice sites and becomes more intricate, as can be seen
in Fig.~\ref{vortexfig}d.
But its phase structure is almost the same as with $\mu=0$.
We should point out that these vortex cells are different from
conventional vortices without optical lattice on a major aspect:
their densities and phases depend on both $r$ and $\theta$.
The normalized power and energy diagrams versus propagation constant
for these vortex cells
at $V_0=1$ are shown in Fig.~\ref{vortexfig}(a).
This figure is similar to that for fundamental solitons [Fig. 1(a)],
except that both $P$ and $E$ here are about four times larger.
Unstable eigenvalues $\sigma$ of vortex-cells are 
determined by simulating the linearized equation (\ref{NLS}) around vortex-cells. 
The results are shown in Fig. \ref{evolution}a for $V_0=1$.
We see that vortex cells experience
an oscillatory instability for $\mu>-7.8$ and become stable for $\mu<-7.8$.
These vortex cells also suffer the VK instability
in the region $\mu > 0.73$ where $dP/d\mu>0$ [see Fig. 3a].
However, the oscillatory instability is
much stronger as it occurs over a wider region and has a higher growth rate.
Fig. ~\ref{vortexfig}(a) shows that vortex cells also have positive energy.
Thus, if a vortex cell is linearly stable, it should be able to resist collapse
under small perturbations \cite{pitaevskii}. However, if it suffers the
linear oscillatory instability discussed above, this instability could
result in power exchange from one part of the
cell to another so that the intensity at some small spots becomes high,
triggering
local collapse to occur. We have observed this scenario numerically. An example
is shown in Fig. ~\ref{evolution}(b,c,d) which displays the development of a
vortex cell with $\mu=0$ and $V_0=1$ when it is amplified by 1\% initially.

We next discuss the effect of varying the potential strength $V_0$ on the
formation and stability of fundamental and vortex solitons.
For this purpose, we have chosen $V_0=1.5$, and repeated most of the above
calculations. The results are summarized as follows:
(i) fundamental solitons and vortex-cells exist at higher values of
$\mu$ (up to approximately 1.35);
(ii) at the same value of $\mu$, both the fundamental and vortex
solitons have lower power ($P$) and higher values of energy ($E$);
(iii) the oscillatory instability suffered by vortex cells is
reduced (it could even be completely suppressed when $V_0$ becomes
even larger). Thus, we conclude that higher lattice potentials stabilize
both fundamental and vortex solitons. 

In conclusion, we have studied new types of fundamental and vortex
solitons in a two-dimensional optical lattice potential, and
shown that both solitons are stable in the strong localization regime.

The work of J.Y. was supported by NSF and NASA.

\end{multicols}
\end{document}